\documentstyle[aps,prl,preprint,amsmath,tighten,epsf]{revtex}
\begin{document}
\draft
\title{Designability of $\alpha$-helical Proteins} 
\author{Eldon G. Emberly, Ned S. Wingreen, Chao Tang$^*$} 
\address{NEC Research Institute, 4 Independence Way, Princeton, NJ
08540, USA \\
$^*$ corresponding author: email: tang@research.nj.nec.com \\
P:~(609)~951-2644 F:~(609)~951-2496}
\maketitle
\date{\today}
\vspace{1in}

{\bf Classification}: Biological, Biophysics

{\bf Manuscript info}: 15 pages, 6 figures, 2 Tables

{\bf Word count}: abstract (120 words), manuscript (3400 words)
\pagebreak
\begin{abstract}
A typical protein structure is a compact packing of connected
$\alpha$-helices and/or $\beta$-strands.  We have developed a method
for generating the ensemble of compact structures a given set of
helices and strands can form.  The method is tested on structures
composed of four $\alpha$-helices connected by short turns.  All such
natural four-helix bundles that are connected by short turns seen in
nature are reproduced to closer than 3.6 Angstroms per residue within
the ensemble. Since structures with no natural counterpart may be
targets for {\it ab initio} structure design, the designability of
each structure in the ensemble -- defined as the number of sequences
with that structure as their lowest energy state -- is evaluated using
a hydrophobic energy.  For the case of four $\alpha$-helices, a small
set of highly designable structures emerges, most of which have an
analog among the known four-helix fold families, however
several novel packings and topologies are identified.
\end{abstract}
\vspace{1in}
\pagebreak
\section{Introduction}
The number of proteins with structures in the Protein Data Bank
continues to grow at an exponential rate. There is a great diversity
of amino-acid sequences in these proteins, yet there is much less
diversity in the structures themselves. Among currently known
structures, only several hundred qualitatively distinct folds have
been identified -- indeed, it has been estimated that there are
only about $1000$ distinct protein folds in
nature\cite{Chothia92,Orengo94,Brenner97,Wang96,Govin99}. Has nature
exhausted all possible folds? If not, how can we design proteins to
adopt folds not seen in nature?

Important progress has been made in designing natural folds ``from
scratch'' \cite{Richardson,Bryson,Kamtekar,Degrado99}. Recently several
attemps have been made to modify natural folds. Dahiyat and
Mayo\cite{Dahiyat97} were able to design a zinc finger that no longer
depended on a zinc ion for stability.  Harbury {\it et
al.}~\cite{Harbury98} were able to design sequences of amino acids so
that the superhelical twist of coiled coils was right handed, in
contrast to the left-handed twist found in nature up to that
time\cite{MacKenzie97}. Kortemme et al. successfully designed a three
stranded $\beta$-sheet protein \cite{Kortemme}.

Combinatorial experimental approaches to creating new protein
structures are also possible.  Studies of the folding of random
amino-acid sequences by Davidson and Sauer \cite{Davidson94}
identified some sequences which appear to fold.  However, the
conformations were not sufficiently rigid to allow structural
determination by either X-ray crystallography or
nuclear-magnetic-resonance techniques to see if there were novel
folds. Recently, Szostak and colleagues \cite{Keefe01} have been able
to find folding proteins by {\it in vitro} evolution. This method can
be used to identify proteins which bind to a particular substrate. It
gives the ability to design for certain function but with no guarantee
that the proteins found in this way will be novel folds. Another
powerful method to evolve for novel functions (or potentially new
folds) is {\it in vitro} DNA recombination \cite{Giver}. But again it
has not been applied to screening for new folds.

Theoretical approaches to the design of qualitatively new folds have
followed two paths: searching within structure space for new folds
\cite{Miller00,Emberly01} and searching in sequence space for
sequences that lead to new folds\cite{Fain01,Simons}. The first
approach has thus far relied on enumerating protein backbones using a
finite set of dihedral-angle pairs\cite{Park95}. In this approach,
enumerating all backbones for proteins of length greater than $30$ is
computationally intractable.  Sampling methods can generate longer
chains, but so far fail to achieve realistic secondary
structures\cite{Emberly01}. The second approach has been attempted
using several schemes. One involves enumerating helical structures
using sequence specific contacts\cite{Fain01}. Another uses a library
of sequences with known structure to assemble possible structures that
a given sequence may adopt\cite{Simons}. However, searching the large
space of sequences for potentially new folds is a huge computational
challenge.

In this paper, we present a computational method for generating
packings of secondary structures which, we believe, will facilitate
the search for novel protein folds and complement the methods
described above.  Our method is motivated by the following
observations: Most naturally occurring protein structures are composed
of two fundamental building blocks, $\alpha$-helices and
$\beta$-strands\cite{Banavar}. A typical protein structure is a
packing of helices and strands connected by turns. The helices and
strands are stabilized by hydrogen bonds, by tertiary interactions and
by the high propensity of some amino acids to form helices and of
others to form strands\cite{Munoz}.  Because some residues are
hydrophobic, the helices and strands pack together in a specific way
to minimize the exposure of the hydrophobic regions to water.

The packing of secondary structural elements, with the connecting
turns cut away, is generally known as a protein's ``stack''.  This
stack, plus information about which elements are connected together by
turns, yields the protein's fold\cite{SCOP}. Our method for generating
protein folds begins by first specifying a fixed number of
$\alpha$-helices and/or $\beta$-strands of fixed lengths, and second
systematically enumerating all of the possible stacks of these
elements.  The great advantage of using fixed secondary structural
elements is that one freezes many of the degrees of freedom of the
chain.  The freezing of these elements can be designed in by choosing
amino acids with appropriate helical or strand propensities. (Loops
can later be used to connect the secondary structures
\cite{Vita99,Liang00}). To test our scheme for generating stacks, we
apply it to the packing of four $\alpha$-helices. Four helix bundles
are a good test case as the natural bundles fall into a small number
of fold families\cite{Harris}, and it has proven possible to design
four-helix bundles through a careful selection of hydrophobobic-polar
sequences\cite{Bryson,Kamtekar}. Our method is able to reproduce the
four-helix-bundle families in the Structural Classification Of
Proteins (SCOP) database\cite{SCOP}.

Within a set of stacks, those with no natural counterparts are
potential candidates for the design of novel protein folds.  To
identify promising candidates, we consider their ``designability".
The designability of a structure is defined as the number of
amino-acid sequences which have that structure as their lowest energy
conformation.  In lattice models, it has been shown that the sequences
associated with highly designable structures have protein-like
properties: mutational stability,\cite{Govin95,Li96} thermodynamic
stability,\cite{Li96,Li98}, fast folding
kinetics\cite{Govin95,Melin99} and tertiary
symmetry\cite{Wang,Wolynes}. Recently, off-lattice studies of protein
structures have also shown that certain backbone configurations are
highly designable, and that the associated sequences have enhanced
mutational and thermodynamic stability\cite{Miller00,Emberly01}.
Hence, we aim to identify those stack configurations with high
designability, and without natural counterparts, as targets for novel
structure design. Several novel four-helix folds are identified.

\section{Results}
We applied our structure generation method (described in detail in
Methods) to the packing of four $\alpha$-helices. We chose each helix
to be 15 residues long\footnote{The procedure was also tested on the
packing of shorter (10 residues) and longer helices (20 residues), with
the short helices producing highly variable packings and the longer
helices tending to always pack into up and down configurations.} (each
helix has a periodicity of $3.6$ residues and a rise of 1.5
Angstroms/residue).  The backbones of turns connecting the helices
were not specified, but the turns were constrained to be short.
Specifically, we discarded a stack if any of the end-to-end distances
between connected helices exceeded $12$ Angstroms. The method
generated a ``complete'' ensemble of four-helix stacks consisting of
$1,297,808$ structures (for a discussion of completeness, see
Methods). This large ensemble of structures was then clustered,
resulting in $188,538$ representative structures.

To test if the method reproduced the natural four-helix bundles, we
selected 11 proteins with short turns, from different SCOP families,
and searched our representative structures for the best fits. To
account for length differences between helices in the SCOP structures
(the lengths ranged from 7-18 residues) and the 15-residue helices in
our model, we chose the shorter length for each comparison.  For the
longer helix of each mismatched pair, we tried all possible
truncations down to the shorter length.  Thus for each pairing of a
SCOP structure with one of our representive structures, we computed
the best fit among all possible combinations of truncations.
Fig.~\ref{fig1} shows four overall best fits among all possible
pairings. For the $11$ natural four-helix bundles, the average crms to
a representative structure was $2.86$ Angstroms.  {Table~\ref{tbl1}
summarizes the results of fitting the natural four-helix bundles to
our representive structures. In all cases, the natural structure had a
counterpart in the representative ensemble at a crms distance of less
than $3.6$ Angstroms per residue.

An important goal is to identify stacks with no natural counterparts
as candidates for the design of novel protein folds.  To identify
which stacks might be promising candidates, we performed a
designability calculation using a hydrophobic energy (see Methods) on
the ensemble of representatives of our four-helix structures.  We used
a random sample of 4,000,000 binary amino-acid
sequences. Fig.~\ref{fig2} shows the results of the designability
calculation. The distribution of designabilities is consistent with
previous results for both lattice\cite{Li96} and off-lattice
models\cite{Miller00,Emberly01} -- namely, there is a small set of
highly designable structures with the great majority of structures
poorly designable or undesignable. The average designability, {\it i.e.}
the average number of sequences per stack, was {4,000,000/188,538 =
21}.  The most designable structure was the lowest energy state of 1813
sequences.

Almost all of the designable structures have an analog amongst the
four-helix fold families. The four most designable distinct folds are
shown in order of designability in Fig.~\ref{fig3}.  The topmost
designable structure is an up-and-down four-helix bundle, the second
most designable fold is a variant of the up-and-down fold except that
there is a crossover connection, the third most designable fold falls
within the $\lambda$-repressor DNA-binding-domain class and the last
fold is an orthogonal array\cite{SCOP}. Table~\ref{tbl2} presents
particular binary sequences which have these structures as lowest
energy folds.  We obtained these sequences by matching them to the
surface area pattern of each of the four folds and then introducing
mutations to maximize the energy gap. The energy gap was defined as
the smallest energy difference to a competing structure at a crms $>$
4 Angstroms ({\it i.e.} a structure with a different fold
type). Sequences were obtained by first calculating the mean
surface-area exposure of each side chain for each structure, and
assigning a hydrophobic residue to each site with surface exposure
below the mean. Point and double mutations were randomly performed on
the sequence by changing H (hydrophobic) to a P (polar) or a P to an
H, and the mutation(s) was kept if the gap was made larger. This
process of mutation was performed until a sequence was obtained where
a mutation at any site made the gap smaller.  The last column in
Table~\ref{tbl2} lists the resulting energy gaps.  Fig.~\ref{fig4}(a)
shows the pattern of surface exposure along each helix for structure
(a) of Fig.~\ref{fig3} along with the corresponding HP pattern (red
for hydrophobic, open for polar).  Notice that the HP pattern of the
optimized sequence in Fig.~\ref{fig4}(a) does not always follow the
rule H at buried site, P at exposed site.  For sites which depart from
the rule, {\it i.e.} a hydrophobic residue on an exposed site, we
found that the nearest competing structure was even more exposed on
that site ({\it e.g.}  site 14 of helix 2 and site 13 of helix 3). For
the site that had a polar residue on a buried site (site 12 of helix
2) the nearest competing structure was less exposed on that site.
Thus, it is sometimes benificial to have hydrophobic residues exposed
and/or polar residues buried in order to ``design-out'' competing
structures\cite{Degrado}.

An important characteristic of natural proteins is their stability
against mutations of individual amino acids.  Generally, it requires
several mutations to cause a natural protein to fail to fold. For our
four most designable distinct folds, we have analyzed the mutational
stability of the optimized sequences (Table~\ref{tbl2}).  We find that
a {\it minimum} of three to five mutations are required to reduce the
energy gap to zero.  For structure (a) of Fig.~\ref{fig3}, four
mutations are required to close the gap; the most effective sites for
these mutations are shown by arrows in Fig.~\ref{fig4}.

Not all of the highly designable structures identified by our method
have close analogs among known natural folds.  We used a vector-based
alignment tool called ``Mammoth''\cite{Ortiz} to align the top 1000
designable structures against 4188 alpha-helical proteins from the
SCOP database. In Fig.~\ref{fig5}, we show two of our designable
structures (left) that had low alignment scores, along with their
closest analogs in the databank (right). The first structure, shown in
Fig.~\ref{fig5}(a), is similar to the POU binding domain. The model
structure was the 15$th$ most designable among the representative
ensemble.  Unlike the POU bundle, which has three helices coiled with
a left-handed twist, the model structure has the same three helices
coiled with a right-handed twist. We found no similar structure with a
right-handed twist in the databank.  The second structure, shown in
Fig.~\ref{fig5}(b), is an orthogonal array, ranking it 80$th$ among
the representative ensemble.  The model structure's closest natural
analog 1AF7 has a long turn connecting helix 1 to helix 2. In the
model fold, helix 1 is reversed allowing it to connect to helix 2 with
a short turn.  These structures, and others with no known natural
counterparts, may be candidates for the design of novel folds.

\section{Discussion and Conclusions}
We have presented a method for generating protein stacks by packing
together fixed secondary structural elements. The method was used to
generate an ensemble of stacks of four $\alpha$-helices. Each of 11
natural structures, stripped of turns, was matched to within 3.6
Angstroms crms by a stack in the model ensemble, despite different
helix lengths in the natural and model structures. The quantitative
similarity between the model structures and the natural four-helix
bundles suggests that the method is a reliable way of exploring the
space of possible stacks.

The designabilities of the generated stacks followed the previously
observed pattern\cite{Li96,Miller00} -- a small set of structures were
highly designable, being lowest energy states of many more than their
share of sequences, while the majority of structures were poorly
designable. The universality of this distribution of designabilities
in model studies suggests that it may apply to real protein structures
as well -- some structures may be intrinsically much more designable
than others. Also consistent with previous model studies, sequences
which fold into highly designable structures were typically 
thermodynamically stable and stable against mutations. We found
that a minimum of 3-5 mutations were required to destabilize
optimized sequences for our most designable structures.
Interestingly, the hydrophobic-polar patterns of these optimized
sequences depart significantly from the simple rule
hydrophobic at buried sites, polar at exposed sites.

Almost all of the most designable four-helix stacks emerging from our
model have analogs among the known four-helix-bundle folds.  This
suggests that nature has found all, or nearly all, designable
four-helix bundles. However, several novel four-helix folds were
identified by our method. These are now the target of design.

\section{Acknowledgments}
The authors would like to thank Jonathan Miller, David Moroz, Ranjan
Mukopadahay, and Chen Zeng for rewarding discussions.

\section{Methods}
{\it Generation of ensemble of stacks} -- The elements of the stack are
chosen depending on the size and type of protein desired. These
elements can be $\alpha$-helices and/or $\beta$-strands. The number of each
type of element is specified, as is  the length in residues
of each element. The sequential arrangement of the elements along the
protein chain is  also specified, along with the maximum
length of the turns connecting elements.

Each element in the stack is assumed to be a rigid body, described by
its center of mass and three Euler angles.  The same simplifying
assumption has been employed previously by Erman, Bahar, and Jernigan
in their work on the packing of pairs of $\alpha$-helices\cite{Erman}.
This method also compliments previous work which has looked at the
packing of fixed secondary structure\cite{Monge1,Monge2,Park2,Chou}.
An element, helix or strand, is specified by its backbone
$\alpha$-carbon-atom positions and its amino-acid side-chain
centroids, the latter taken to lie in the direction of the
$\beta$-carbon at a distance of 2.1 Angstroms from the
$\alpha$-carbon. Helices are constructed using a helical periodicity
of 3.6 residues and a helical rise of 1.5 Angstroms/residue. Strands
are created by using a single backbone dihedral angle pair from the
beta-strand region of the Ramachandran plot. A stack is generated by
first randomly selecting the center of mass and Euler angles for each
element (if an element's center of mass and angles cause it to violate
self-avoidance with one of the other elements, then its degrees of
freedom are re-selected randomly).  Then these variables are relaxed
so as to minimize the packing energy (described in detail below). A
local minimum of the packing energy is found using a
conjugate-gradient method, described in Numerical
Recipes\cite{Press92}. This yields a stack.  With the centers of mass
and angles determined, various symmetry operations are then performed
to generate additional stacks. For $\alpha$-helical elements these are
screw operations which correspond to rotating the helix by $\pm 100$
degrees and translating it by $\pm 1.5$ Angstroms along the helix
direction.  For $\beta$-strands, slide operations correspond to
translating each residue up or down by one residue along the strand
direction.  Each stack is then checked to see if it satisfies a set of
supplied constraints. For instance, stacks that exceed a specified
total surface exposure or compactness measure, or have end-to-end
distances of connected elements which exceed some cut-off, are
excluded from the set. If a stack satisfies the constraints, it is
added to the ensemble.  Stacks are generated in this way until the
ensemble of possible stacks for this model is complete, as discussed
below.

The choice of packing energy $E_{\rm packing}$ is motivated by the
hydrophobic force, which produces the compact stacks found in nature.
The first term of the packing energy is
\begin{equation}	
	E_1 = \sum_i s_i
\end{equation}
where $s_i$ is the surface exposure to water of the $i^{th}$ residue along
the chain. The surface exposure of each residue is calculated 
by approximating each side-chain as a sphere with radius 
$R_S = 3.1$ Angstroms centered at a distance $L = 2.1$ Angstroms from 
its $\alpha$-carbon atom, in the direction of the $\beta$-carbon. 
The surface exposure $s_i$ of each side-chain sphere is found 
using the method of Flower \cite{Flower97}, with a water molecule 
represented as a sphere of radius $R_{H_2O} = 1.4$ Angstroms.

We add to this hydrophobic energy a second term which represents the
effect of excluded volume. This term $E_2$ is a pairwise repulsive
energy among backbone $\alpha$-carbon atoms and side-chain centroids
on {\it different elements}. The excluded volume energy is given by,
\begin{equation}
       E_2 =  V_0 \sum \Bigl[
	\Bigl(\frac{2 R_{{\alpha}}}{r^{\alpha}_{i,j}}\Bigr)^{12} + 
		\Bigl(\frac{2 R_{{\beta}}}{r^{\beta}_{i,j}}\Bigr)^{12}  + 
	\Bigl(\frac{R_{{\alpha}}+R_{{\beta}}}
              {r^{\alpha,\beta}_{i,j}}\Bigr)^{12} \Bigr]
\end{equation}
where $R_{{\alpha}} = 1.75$ Angstroms and $R_{{\beta}} = 2.25$
Angstroms are sphere sizes for the backbone $\alpha$-carbon atoms and
side-chain centroids, respectively, 
$r^{\alpha}_{i,j}$ is the distance between
backbone $\alpha$-carbon atoms $i$ and $j$, $r^{\beta}_{i,j}$ is the
distance between centroids $i$ and $j$, and $r^{\alpha,\beta}_{i,j}$
is the distance between backbone $\alpha$-carbon atom $i$ and centroid
$j$. $V_0$ sets the scale of the repulsive energy.

Lastly, we include a weak compression energy $E_3$ and
an energy $E_4$ due to tethers between the ends of connected
elements. These energies have the form,
\begin{equation}
	  E_3 =  \frac{K}{2} r_g^2,
\end{equation}
where $r_g$ is the radius of gyration of the entire
stack\footnote{$r_g^2 = 1/N \sum_j 
({\bf R}^{stack}_{CM} - {\bf r}^{\beta}_j)^2$
where ${\bf R}^{stack}_{CM}$ is the center of mass of the entire stack and
${\bf r}^{\beta}_j$ is the position of centroid $j$.}, and
\begin{equation}
          E_4 = \sum_i \frac{K_T}{2}(d_{i,j} - d^{\, 0}_{i,j})^2
          \theta(d_{i,j}-d_{i,j}^{\, 0}),
\end{equation}
where, $d_{i,j}$ is the distance between the connected ends of
tethered elements $i$ and $j$, and $d^{\, 0}_{i,j}$ is a specified
equilibrium length (for the case of the helices above we used $12$
Angstroms) and $\theta$ is a step function that is $0$ if $d_{i,j} <
d_{i,j}^{\, 0}$ and $1$ otherwise.  The spring constants, $K$ and
$K_T$ are chosen to be small so that these terms act as weak
perturbations.

The actual minimization of the total energy $E_{\rm packing} =
E_1+E_2+E_3+E_4$ using the conjugate-gradient method proceeds in
steps, akin to annealing. The scheduled parameter is $V_0$. Initially
$V_0$ is chosen to be large, so that there is a large repulsion
between all the elements. (The starting value of $V_0$ varies
depending on the number and size of the chosen elements.  The initial
$V_0$ is chosen so as to generate a smooth collapse of the elements.
For the case of four-helix bundles we chose a starting $V_0$ of 35.0)
At a given $V_0$, a minimum of $E_{\rm packing}$ is found for the full
set of center of mass and angle variables. $V_0$ is then reduced by a
constant factor (90 \%) and a small random change is made to each
degree of freedom. (The size of the random ``kick'' is also scaled
along with $V_0$, with the initial kick being $1$ Angstrom for the
centers of mass and $15$ degrees for each Euler angle). The $V_0$
schedule is terminated when any two centroids are at a distance less
than some specified contact distance, taken to be $2 R_S$. At this
point, $E_3$ and $E_4$ are set to zero, leaving only $E_1$ and $E_2$
to be minimized in the last conjugate gradient step.  $V_0$ is then
set to its final value\footnote{The final value of $V_0$ is determined
by a fitting procedure involving a naturally occurring stack composed
of similar elements.  Specifically, $V_0$ is chosen to minimize the
crms distance between the stack before and after a conjugate-gradient
minimization, with fixed $E_1$ and $E_2$. For the four helix bundle we
found that a value of $V_0 = 0.05$ produced the best fits to the
chosen SCOP structures. $V_0$ controls the inter-helical separation,
and thus changing it by a few percent only serves to increase or
decrease the contact distances of helices. Making $V_0$ signficantly
different from this makes the sidechain spheres unphysically small or
large, which can lead to unreasonable packings.}  and the last
conjugate-gradient minimization is performed to yield final values of
each rigid element's center of mass and orientation angles.

{\it Flexible elements} -- The method described above can be
generalized to allow flexibility of the secondary structural elements.
In natural protein structures, $\alpha$-helices are relatively rigid,
while $\beta$-strands are more flexible. Hence, the extension of the
method to include flexible elements is more important in the case of
$\beta$-strands.

The flexural modes of rod shaped objects are bending, stretching, and
twisting. All these internal flexural modes can be included in the
generation of stacks for both $\alpha$-helices and $\beta$-strands.
It is possible to determine the appropriate degree of flexibility for
each internal mode by reference to known protein structures.  A
harmonic energy function $E_{\rm flex}$ for these flexural modes can
then be added to the packing energy, with coefficients chosen to
reproduce the degree of flexibility observed in natural proteins. For
example, if the degree of bending of an $\alpha$-helix is represented
by the angle $\theta$, then the additional term in $E_{\rm packing}$
representing this mode would be
\begin{equation}	
        E_{\theta} = \frac{c_{\theta}}{2} \, \theta^2,
\end{equation}
where the constant $c_{\theta}$ can be chosen so that the average
degree of bending $\langle\theta^2\rangle$ in the generated stacks
matches that observed in natural structures.  In the current work,
however, we focus on $\alpha$-helical proteins and only rigid
elements are considered.

{\it Hydrogen bonding} --
In natural proteins, $\beta$-strands are typically stabilized by the
formation of hydrogen bonds between strands.  To generate stack
configurations which include strands it is therefore important
to include an inter-strand hydrogen-bonding energy $E_{hb}$ in the
packing energy $E_{\rm packing}$. One form of a hydrogen-bonding
energy function is given in \cite{Dahiyat2}.

{\it Completeness of stack ensemble} -- Designability is determined
via a competition for amino-acid sequences within a complete set of
stacks. Since the method for generating stacks is based on random
sampling, a criterion must be specified for when to stop sampling. We
stop the generation of structures when a specified fraction of newly
generated stacks already occurs in the previously generated
ensemble. If the fraction is not satisfied, the newly generated
structures are added to the ensemble, and more stacks are randomly
generated. We use crms to measure simlarity between the ensemble and
the newly generated structures, and consider two structures to be
similar if their crms is less than 1.5 Angstroms.
The distance measure, $\mathrm{crms}$, is defined as
\begin{equation}
	 (\mathrm{crms})^2 = \frac{1}{N} \sum_i \,(\vec{r}^{\, s}_i -
\vec{r}^{\, s'}_i)^2
\end{equation}
where $\vec{r}^{\, s/s'}_i$ is the position of the $i^{th}$
$\alpha$-carbon for the $s/s'$ stack and $N$ is the number of backbone
$\alpha$-carbons.  The stacks $s$ and $s'$ are aligned by performing
a least-squares fit using crms as the metric.  We demand that 95\%
of the newly generated structures be similar to one of the structures
in the ensemble before stopping the structure generation procedure.

{\it Clustering ---} Many of the randomly generated stacks form
clusters of closely related structures. It is computationally 
advantageous to reduce the sample by retaining only one member of
each cluster. These representative structures
are selected in the following way. The entire set of stacks is  
sorted according to total surface exposure, {\it i.e.} from
most compact to least compact. Starting at the top of this list with 
the most compact stack, we eliminate all stacks that are closer
to it than $1.5$ Angstroms crms.  This
process is repeated for the next most compact structure in the list
until the end of the list is reached. We can typically compress the
large ensemble of structures by a factor of $3-5$ in this way.

{\it Designabilities of stacks} -- The designabilities of the 
representative stacks, after clustering, are determined
by allowing the structures to compete for a random sample of possible
amino-acid sequences. The ``designability" of a stack is defined as
the number of sequences for which that stack has the lowest energy.
We assume that the hydrophobic energy is the dominant term
contributing to the energy of a sequence on a given structure. 
This energy is given by
\begin{equation}
		E_{h} =  \sum_i h_i s_i, \label{eqn:hydro}
\end{equation}
where $h_i$ is the hydrophobicity of the $i^{th}$ element of the
sequence and $s_i$ is the fractional surface exposure of the $i^{th}$
side-chain sphere in the particular stack.  For each sequence
considered, the lowest energy stack in the representative ensemble is
determined.  By sampling a large number of randomly selected
sequences, it is possible to reliably estimate the relative
designabilities of different stacks.

For the designability calculation, we employed binary sequences
consisting of only two types of amino acids. Such sequences are also
known as ``HP-sequences" for hydrophobic (H) and polar (P) amino
acids. In previous studies, we found only minimal differences in the
designabilities of top structures when binary sequences and sequences
with a continuous distribution of hydrophobicities were used
\cite{Miller00}.  The two hydrophobicity values can be written as $h_i
= h_0 \pm \delta h$, where $h_0$ is a compactification energy, and
$\delta h$ measures the relative difference between hydrophobic and
polar residues. From the Miyazawa-Jernigan matrix \cite{Miyazawa} of
amino-acid interaction energies, we infer a typical energy difference
between hydrophobic and polar residues of $1.5 k_BT$/contact.  On
average a buried residue makes four non-covalent contacts, therefore
we take $2\delta h = 6.0 k_BT$. The compactification energy $h_0$ was
determined by fitting the surface-area distribution of the set of 11
natural four-helix bundles given in Results to the surface-area
distributions for the 100 most designable four-helix stacks, using
different values of $h_0$ to assess designability. The best fit is
shown in Fig.~\ref{figA}, and this corresponded to $h_0 = 2 k_BT$ .
Thus in our model hydrophobic residues have a hydrophobicity of $5 k_B
T$ and polar residues $-1 k_B T$.

If flexible $\alpha$-helices and/or $\beta$-strands are
employed in generating stacks, the energy $E_{\rm flex}$ associated with the
flexural modes can be added to the hydrophobic energy $E_{h}$.  
Similarly, if inter-strand hydrogen bonding is included,
$E_{hb}$ can be added as well. The energies $E_{\rm flex}$ and
$E_{hb}$ add a sequence independent contribution
to each stack.

%FIGURES
\begin{table}[!t]
\begin{tabular}{|| c | c ||} \hline
PDB ID  & crms (Angstroms) \\ \hline
1FLX &  2.96  \\ \hline
1FFH &  3.54  \\ \hline
1E6I &  2.85  \\ \hline
1CB1 &  1.65  \\ \hline
1CEI &  2.95  \\ \hline
1A24 &  2.85  \\ \hline
1POU &  2.81  \\ \hline
1AU7 &  3.02  \\ \hline
1EH2 &  2.74  \\ \hline
1IMQ &  2.75  \\ \hline
1DNY &  3.44  \\ \hline
\end{tabular}
\caption{Results of fitting selected set of $11$ proteins from SCOP
database to ensemble of model four-helix bundles.}\label{tbl1}
\end{table}

\begin{table}[!t]
\begin{tabular}{||c|c|c|c||} \hline
Structure  & Sequence & Energy Gap ($k_BT$)  & Minimum Mutations \\ \hline
a  helix 1 &             PPHHHHHHPHHPPHH  & &\\ 
a  helix 2 &             HHPPHHPHHPHPHHP  & &\\ 
a  helix 3 &             PPHHPPHHPHHHHHH  &   6.65  & 4 \\
a  helix 4 &             PHHPPHHPHHPHPHP  & & \\  \hline

b  helix 1 &             HHPPHHPHHPHHHHP & & \\ 
b  helix 2 &             HHHPPHHPHHHPHHP & &\\ 
b  helix 3 &             PPHHHHHPHPPPHHP &   5.85 & 3 \\ 
b  heilx 4 &             HPHHHPHHPHHPHHH & & \\  \hline

c  helix 1 &             HHHHPPHPPPHPPHP & & \\ 
c  helix 2 &             PHPPHHPPPHPPHHP & & \\ 
c  helix 3 &             PHPPHHHPPHHPPPP &   8.3 &  5 \\ 
c  helix 4 &             PPPPHPPPHPPHHHH & & \\  \hline

d  helix 1 &             HPHHHPHHPPHHHPP & & \\ 
d  helix 2 &             PHHPHHHPHHPPPHP & & \\ 
d  helix 3 &             PHHPHHHPHHPHHPP &   4.70 & 3\\ 
d  helix 4 &             PHHHPHHPHHHHHHH & & \\ \hline 
\end{tabular}
\caption{Results for the four most designable distinct folds for the
model four-helix bundles shown in Fig.~\ref{fig3}. Column 2 gives the
optimized hydrophobic-polar patterning of each of the length 15
helices.  For these sequences, the third column gives the energy gap
in $k_BT$ to the nearest distinct structural competitor. The last
column gives the minimum number of point mutations necessary to reduce
the energy gap to zero.}\label{tbl2}
\end{table}

\begin{figure}[!t]
\begin{center}
\caption{Representative fits for four SCOP proteins (left column) to model
four-helix bundles (right column): (a) fit for 1EH2 (crms = 2.74 
Angstroms/residue),
(b) fit for 1FFH (crms = 3.54 Angstroms/residue), (c) fit for 1CEI
(crms = 2.95 Angstroms/residue), (d) fit for 1POU (crms = 2.81
Angstroms/residue). Numbers indicate helix number and their location
indicates the beginning of the given helix.}
\label{fig1}
\end{center}
\end{figure} 

\begin{figure}[!t]
\begin{center}
\caption{Histogram of the number of structures with a given
designability for the representative structures of the four-helix-bundle
ensemble. Only a few of the structures are highly designable,
{\it i.e.} are lowest energy states of a large number of sequences.
Most structures are lowest energy states of few or no sequences.}
\label{fig2}
\end{center}
\end{figure} 

\begin{figure}[!t]
\begin{center}
\caption{Four most designable distinct four-helix folds: (a)
up-and-down fold, (b) up-and-down with a cross-over connection fold,
(c) $\lambda$-repressor-type fold, (d) orthogonal-array fold. Numbers
indicate helix number and their location indicates the beginning of the
given helix.}
\label{fig3}
\end{center}
\end{figure} 

\begin{figure}[!t]
\begin{center}
\caption{Surface-area exposure for each of the four helices for
structure (a) in Fig.~\ref{fig3} colored with the hydrophobic-polar
pattern of the optimized sequence (red bar = hydrophobic, open bar =
polar). All sites with $<$ 10\% exposure are occupied by hydrophobic
amino acids.  Also shown are the four mutation sites (arrows) which
reduce the energy gap between this structure and its competitor to
zero (site 2, 6 and 15 of helix 2 and site 3 of helix 3).}
\label{fig4}
\end{center}
\end{figure} 

\begin{figure}[!t]
\begin{center}
\caption{Two designable four-helix folds with no known natural
analogs.  On the right are the closest aligned naturally occuring
folds [35], and on the left are the model structures. (a) 1POU
has a left-handed twist of the top three helices. The model structure
has a right-handed twist of these helices. (b) 1AF7 has a long turn
connecting helix 1 to helix 2. The model structure has helix 1
reversed, allowing a short turn between helix 1 and helix 2.}
\label{fig5}
\end{center}
\end{figure} 

\begin{figure}[!t]
\begin{center}
\caption{Best fit of surface distribution of the $11$ SCOP proteins to
top $100$ designable structures found using $h_0 = 2 k_BT$.}
\label{figA}
\end{center}
\end{figure}

\newpage
\begin{figure}               
\vspace{6cm}                 
\centerline{\epsfxsize=12cm 
\epsffile{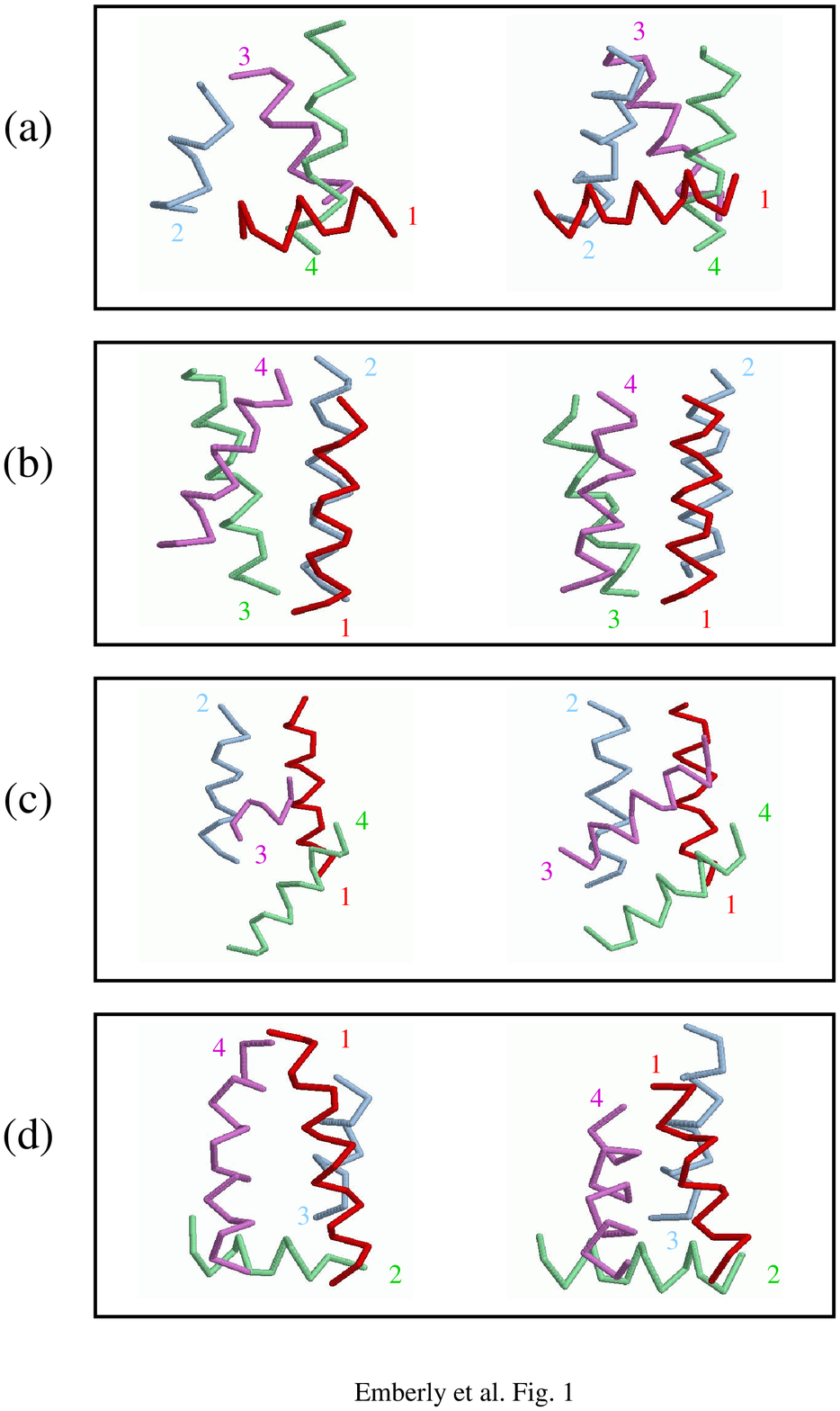}}      
\end{figure}                 
%%%%%%%%%%%%%%               

\newpage
\begin{figure}               
\vspace{6cm}                 
\centerline{\epsfxsize=12cm 
\epsffile{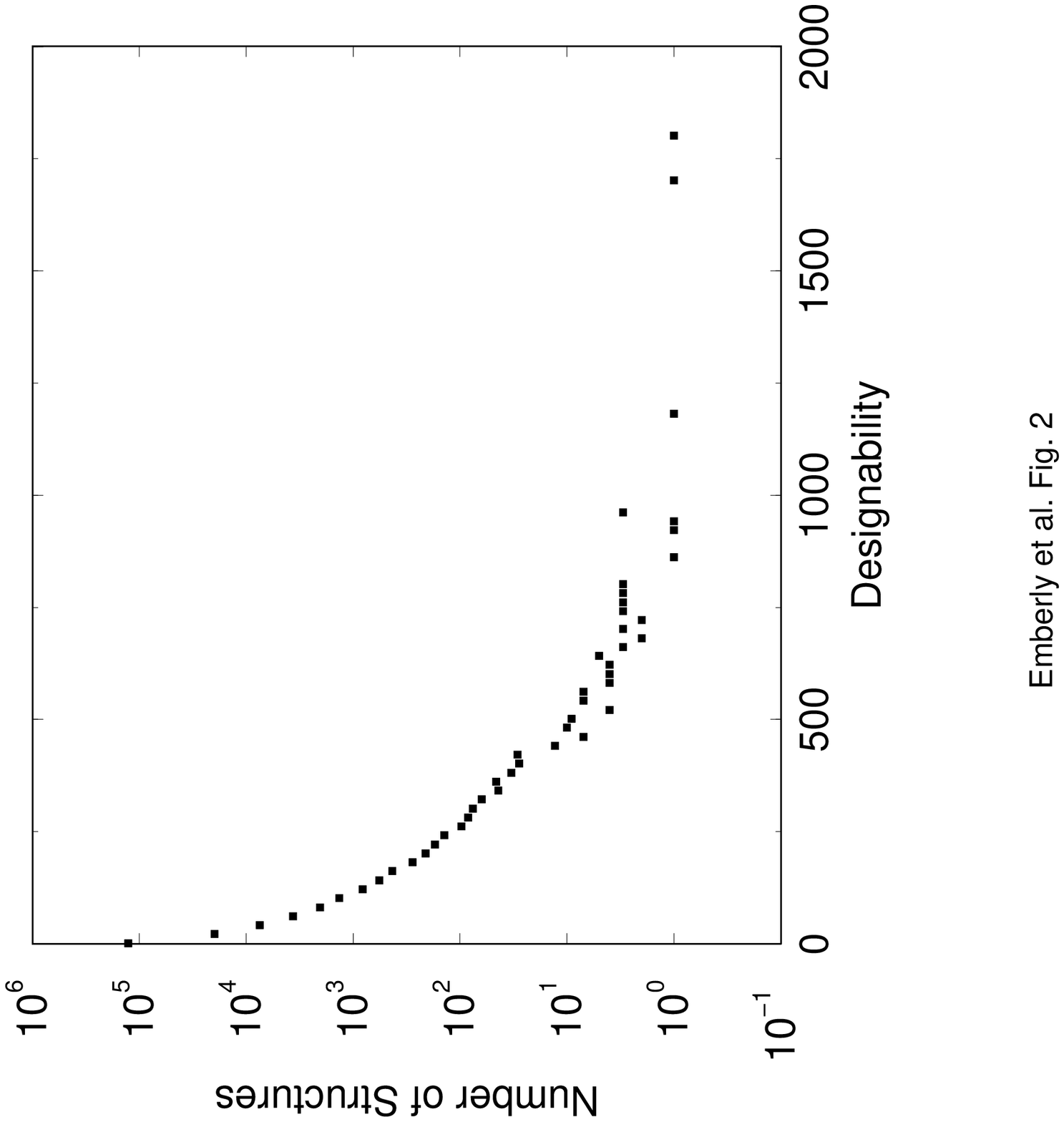}}      
\end{figure}                 
%%%%%%%%%%%%%%               

\newpage
\begin{figure}               
\vspace{6cm}                 
\centerline{\epsfxsize=12cm 
\epsffile{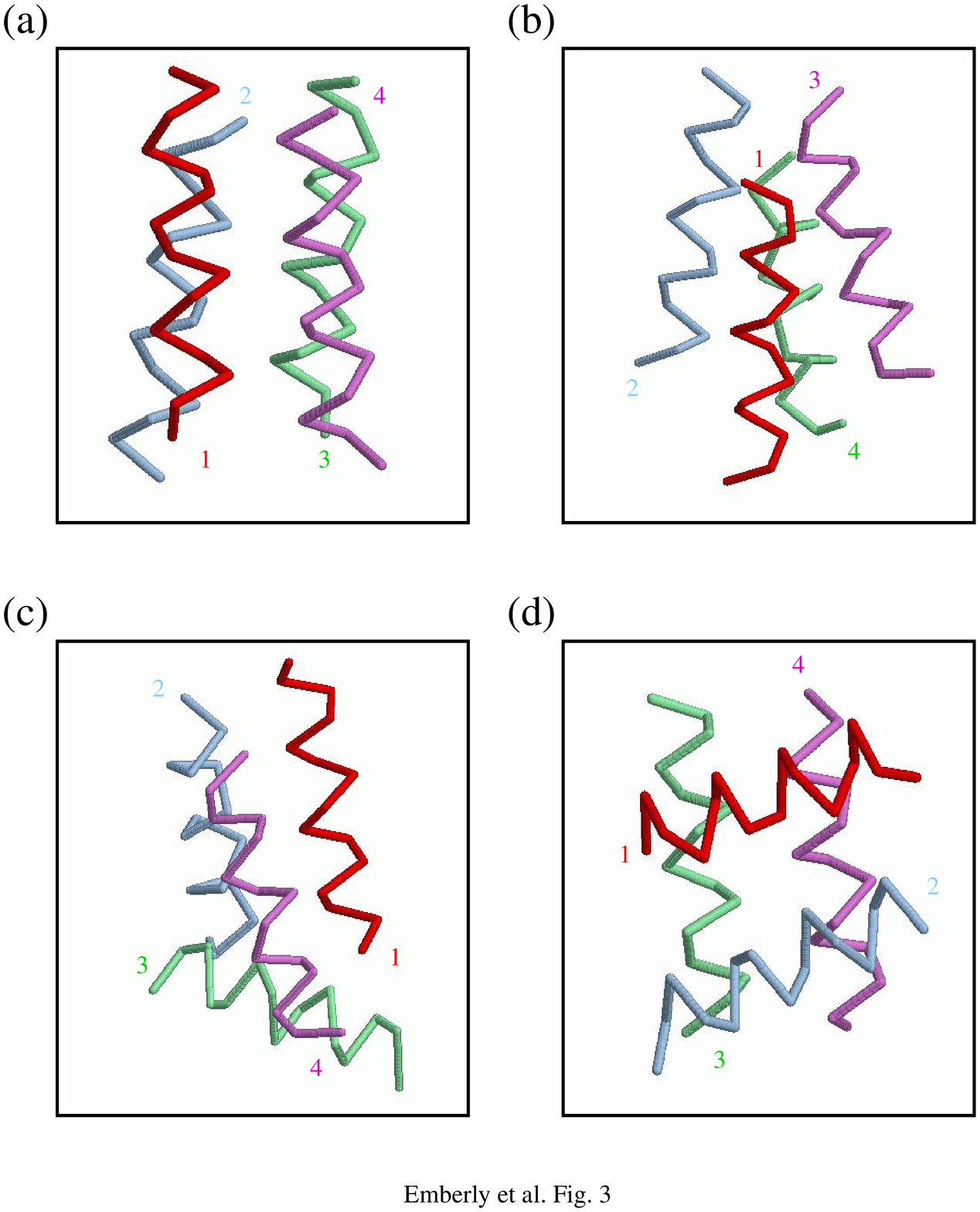}}      
\end{figure}                 
%%%%%%%%%%%%%%               

\newpage
\begin{figure}               
\vspace{6cm}                 
\centerline{\epsfxsize=12cm 
\epsffile{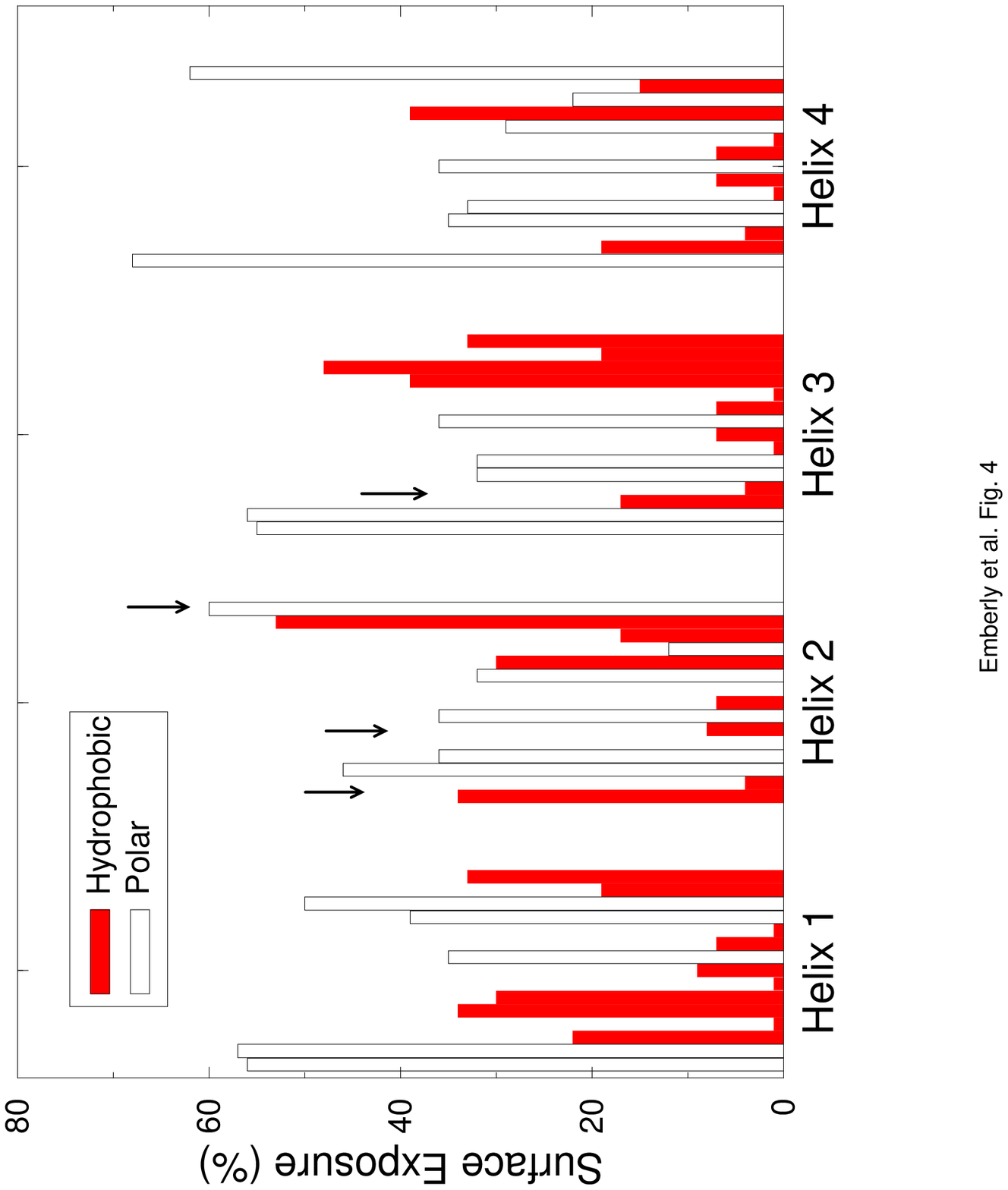}}      
\end{figure}                 
%%%%%%%%%%%%%%               

\newpage
\begin{figure}               
\vspace{6cm}                 
\centerline{\epsfxsize=12cm 
\epsffile{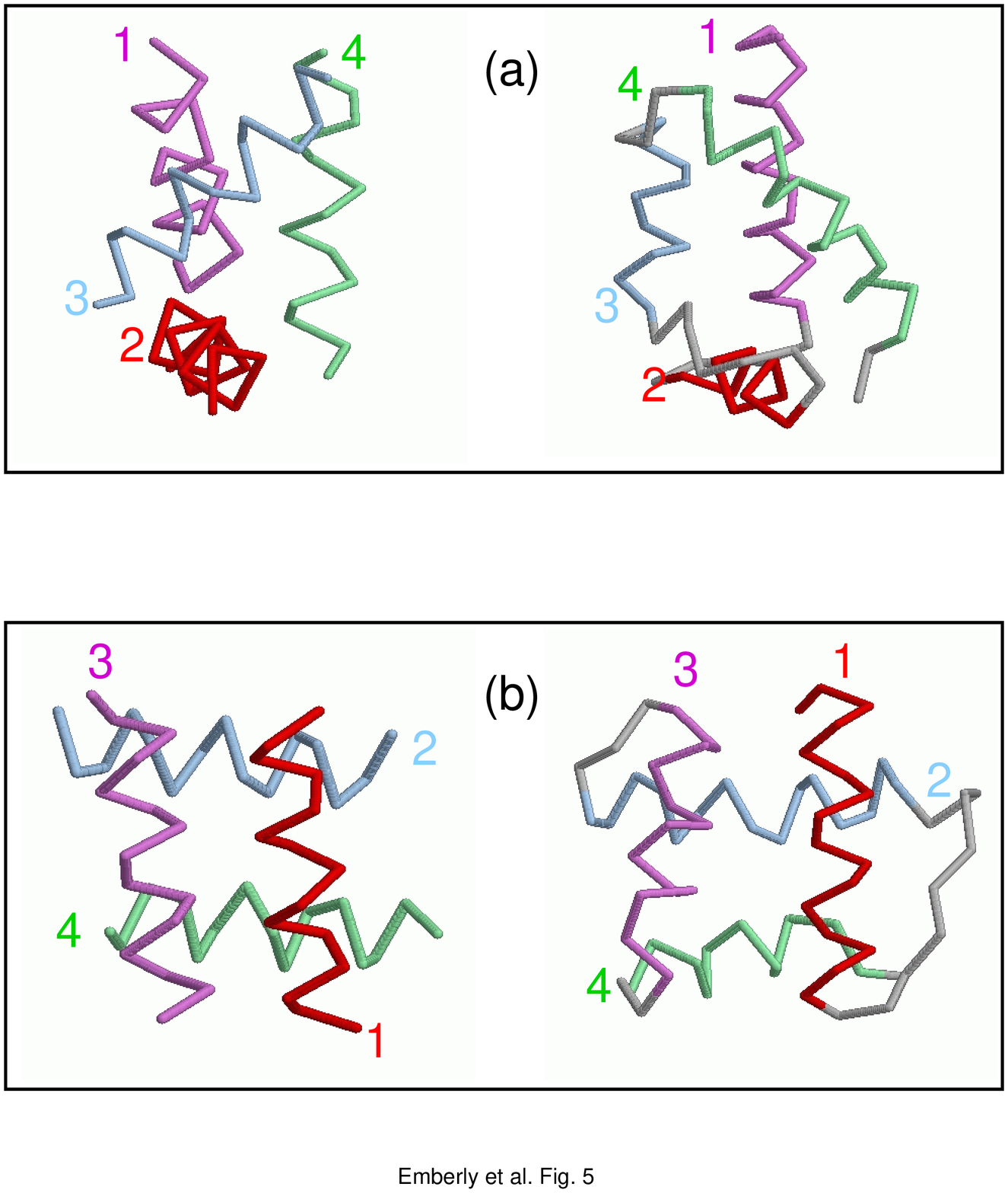}}      
\end{figure}                 
%%%%%%%%%%%%%%               

\newpage
\begin{figure}               
\vspace{6cm}                 
\centerline{\epsfxsize=12cm 
\epsffile{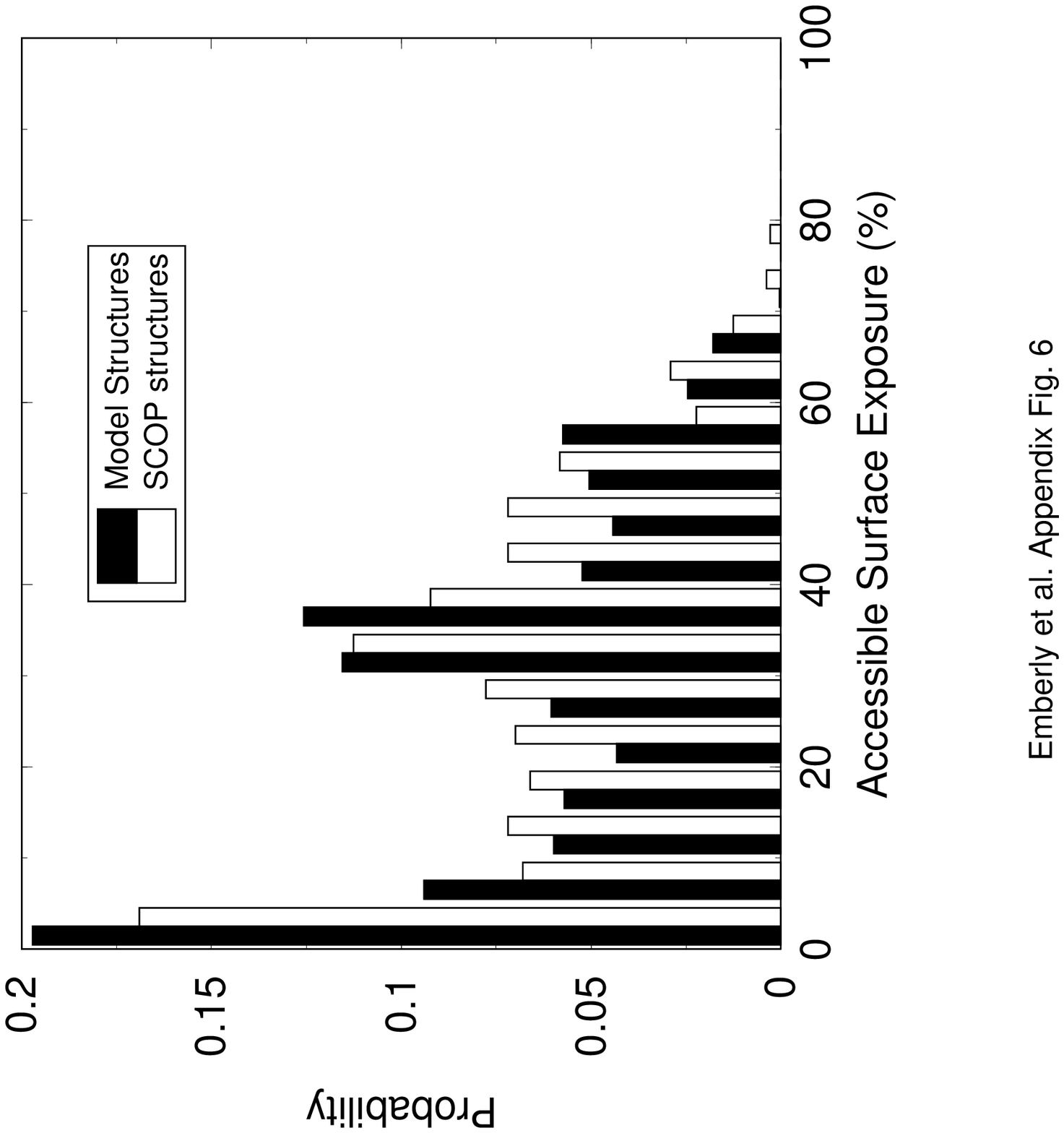}}      
\end{figure}                 
%%%%%%%%%%%%%%               

\end{document}